\documentclass[preprint, 12pt]{elsarticle}

\usepackage{listings}

\usepackage{graphicx}
\usepackage{amssymb}

\biboptions{comma,square}

\def\apgt{\ {\raise-.5ex\hbox{$\buildrel>\over\sim$}}\ }
\def\aplt{\ {\raise-.5ex\hbox{$\buildrel<\over\sim$}}\ }
\def\lt{\ {\raise-.5ex\hbox{$\buildrel>$}}\ }
\def\gt{\ {\raise-.5ex\hbox{$\buildrel<$}}\ }

\journal{Journal of Computational Physics}

\begin{document}

\begin{frontmatter}


\title{A sparse octree gravitational N-body code that runs entirely on
  the GPU processor}


\author[aut1]{Jeroen B\'{e}dorf}\corref{cor1}
\ead{bedorf@strw.leidenuniv.nl}
\author[aut2,aut1]{Evghenii Gaburov}
\author[aut1]{Simon Portegies Zwart}

\cortext[cor1]{Corresponding author}

\address[aut1]{Leiden Observatory, Leiden University, P.O. Box 9513, 2300 RA Leiden, 
The Netherlands
}
\address[aut2]{Northwestern University, 2131 Tech Drive, Evanston 60208, IL, USA
}

\begin{abstract}
We present the implementation and performance of 
a new gravitational $N$-body tree-code that is specifically designed
for the graphics processing unit (GPU)\footnote{The code is publicly available at: \newline {\tt
      http://castle.strw.leidenuniv.nl/software.html}} .
All parts of the tree-code algorithm are executed on the GPU. We present algorithms
for parallel construction and traversing of sparse octrees. These algorithms 
are implemented in CUDA and tested on NVIDIA GPUs, but they are portable 
to OpenCL and can easily be used on many-core devices from other manufacturers. 
This portability is achieved by using general parallel-scan and sort methods. 
The gravitational tree-code outperforms tuned CPU code
during the tree-construction and shows a performance improvement of
more than a factor 20 overall, resulting in a processing rate
of more than $2.8$ million particles per second.
\end{abstract}


\end{frontmatter}

\section{Introduction}\label{sect:introduction}

A common way to partition a three-dimensional domain is the use of
octrees, which recursively subdivide space into eight octants. This
structure is the three-dimensional extension of a binary tree, which
recursively divides the one dimensional domain in halves. One can
distinguish two types of octrees, namely dense and sparse. In the
former, all branches contain an equal number of children and the
structure contains no empty branches.  A sparse octree is an octree of
which most of the nodes are empty (like a sparse matrix), and the
structure is based on the underlying particle distribution. In this
paper we will only focus on sparse octrees which are quite typical for
non-homogenous particle distributions.

Octrees are commonly used in applications that require distance or
intersection based criteria. For example, the octree data-structure
can be used for the range search method \cite{compgeom:2000}. On a set
of $N$ particles a range search using an octree reduces the complexity
of the search from ${\cal O}(N)$ to ${\cal O}(\log N)$ per
particle. The former, though computationally expensive, can easily be
implemented in parallel for many particles. The later requires more
communication and book keeping when developing a parallel
method. Still for large number of particles ($\sim N \geq 10^ 5$)
hierarchical\footnote{ Tree data-structures are commonly referred to
  as hierarchical data-structures} methods are more efficient than
brute force methods. Currently parallel octree implementations are found in a wide
range of problems, among which self gravity simulations, smoothed
particle hydrodynamics, molecular dynamics, clump finding, ray tracing
and voxel rendering; in addition to the octree data-structure
these problems often require long computation times.  For high
resolution simulations ($\sim N\geq10^5$) 1 (Central Processing Unit)
CPU is not sufficient. Therefore one has to use computer clusters or
even supercomputers, both of which are expensive and scarce. An
attractive alternative is a Graphics Processing Unit (GPU).

Over the years GPUs have grown from game devices into more general
purpose compute hardware. With the GPUs becoming less specialised,
new programming languages like Brook \cite{1015800}, CUDA
\cite{CUDAGuide3.2} and OpenCL \cite{OpenCL} were introduced and allow
the GPU to be used efficiently for non-graphics related problems. One
has to use these special programming languages in order to be able to
get the most performance out of a GPU. This can be realized by
considering the enormous difference between today’s CPU and GPU. The
former has up to 8 cores which can execute two threads each, whereas a
modern GPU exhibits hundreds of cores and can execute thousands of
threads in parallel.  The GPU can contain a large number of cores,
because it has fewer resources allocated to control logic compared to
a general purpose CPU. This limited control logic renders the GPU
unsuitable for non-parallel problems, but makes it more than an order
of magnitude faster than the CPU on massively parallel problems
\cite{CUDAGuide3.2}. With the recent introduction of fast double
precision floating point operations, L1 and L2 caches and ECC memory
the GPU has become a major component in the High Performance Computing
market.  The number of dedicated GPU clusters is steadily increasing
and the latest generation of supercomputers have nodes equipped with
GPUs, and have established themselves in the upper regions of the
top500\footnote{Top500 Supercomputing November 2010 list,
  http://www.top500.org}.

This wide spread of GPUs can also be seen in the acceptance of GPUs in
computational astrophysics research. For algorithms with few data
dependencies, such as direct $N$-body simulations, programming the GPU
is relatively straightforward.  Here various implementations are able
to reach almost peak-performance \cite{2007NewA...12..641P,
  2007astro.ph..3100H, 2008NewA...13..103B} and with the introduction
of $N$-body libraries the GPU has taken over the GRAPE (GRAvity PipE
\cite{1998sssp.book.....M}) \footnote{ The GRAPE is a plug-in board
  equipped with a processor that has the gravitational equations
  programmed in hardware.}  as preferred computation device for stellar
dynamics \cite{Gaburov2009630}. Although it is not a trivial task to
efficiently utilise the computational power of GPUs, the success with
direct $N$-body methods shows the potential of the GPU in practice.
For algorithms with many data dependencies
or limited parallelism it is much harder to make efficient use of
parallel architectures.  A good example of this are the gravitational
tree-code algorithms which were introduced in 1986
\cite{1986Natur.324..446B} as a sequential algorithm and later
extended to make efficient use of vector machines
\cite{1990JCoPh..87..161B}.  Around this time the GRAPE hardware was
introduced which made is possible to execute direct $N$-body
simulations at the same speed as a simulation with a tree-code
implementation, while the former scales as ${\cal O}(N^2)$ and the
latter as ${\cal O}(N\log N)$. The hierarchical nature of the
tree-code method makes it difficult to parallelise the algorithms, but
it is possible to speed-up the computational most intensive part,
namely the computation of gravitational interactions. The GRAPE
hardware, although unsuitable for constructing and traversing the
tree-structure, is able to efficiently compute the gravitational
interactions. Therefore a method was developed to create lists of
interacting particles on the host and then let the GRAPE solve the
gravitational interactions \cite{1991PASJ...43..841F,
  2004PASJ...56..521M}.  Recently this method has successfully been
applied to GPUs \cite{citeulike:4604010, 1654123, Hamada:2010:TAN:1884643.1884644}. 
With the GPU being
able to efficiently calculate the force interactions, other parts like
the tree-construction and tree-traverse become the bottleneck of the
application. Moving the data intensive tree-traverse to the GPU
partially lifts this bottleneck \cite{OctGravICCS10, YokotaBarba}.  This method
turns out to be effective for shared time-step integration algorithms,
but is less effective for block time-step implementations. In a block
time-step algorithm not all particles are updated at the same
simulation time-step, but only when required. This results in a more
accurate (less round-off errors, because the reduced number of
interactions) and more efficient (less unnecessary time-steps)
simulation.  The number of particles being integrated per step can be
a fraction of the total number of particles which significantly
reduces the amount of parallelism. Also the percentage of time spent
on solving gravitational interactions goes down and other parts of the
algorithm (e.g. construction, traversal and time integration) become
more important.  This makes the hierarchical tree $N$-body codes less
attractive, since CPU-GPU communication and tree-construction will
become the major bottlenecks \cite{2008NewA...13..103B,
  OctGravICCS10}.  One solution is to implement the tree-construction
on the GPU as has been done for surface reconstruction
\cite{Zhou.Surface} and the creation of bounding volume hierarchies
\cite{LauterbachBVH, Pantaleoni:2010:HHL:1921479.1921493}. An other
possibility is is to implement all parts of the algorithm on the GPU using 
atomic operations and particle insertions \cite{Burtscher}
here the authors, like us, execute all parts of the algorithm on the GPU.
When we were in the final stages of finishing the paper we were able to 
test the implementation by Burtscher et al. \cite{Burtscher}. It is 
difficult to compare the codes since they have different monopole
expansions and multipole acceptance criteria (see Sections
\ref{Sect:NodeProp} and \ref{Sect:OpeningAngle}). However, even though our implementation
has higher multipole moments (quadrupole versus monopole) and a more 
strict multipole acceptance criteria it is at least 4 times faster.

In this work we devised algorithms to execute the tree-construction on
the GPU instead of on the CPU as is customarily done. In addition we
redesign the tree-traverse algorithms for effective execution on the
GPU. The time integration of
the particles is also executed on the GPU in order to remove the
necessity of transferring data between the host and the GPU
completely. This combination of algorithms is excellently suitable for
shared and block time-step simulations. Although here implemented as
part of a gravitational $N$-body code (called {\tt Bonsai}, 
Section \ref{Sect:GravTree}), the algorithms are applicable
and extendable to related methods that use hierarchical data
structures.

\section{Sparse octrees on GPUs}

The tree construction and the tree-traverse rely on scan algorithms,
which can be efficiently implemented on GPUs. (\ref{Sect:Scan}).  Here
we discuss the main algorithms that can be found in all hierarchical
methods.  Starting with the construction of the tree-structure in
Section~ \ref{Sect:TreeConstruction}, followed by the method to
traverse the previously built tree-structure in
Section~\ref{Sect:TreeTrav}. The methods that are more specific for a
gravitational $N$-body tree-algorithm are presented in Section
\ref{Sect:GravTree}.

\subsection{Tree construction}\label{Sect:TreeConstruction}
The common algorithm to construct an octree is based on sequential
particle insertion \cite{1986Natur.324..446B} and is in this form not
suitable for massively parallel processors. However, a substantial
degree of parallelism can be obtained if the tree is constructed
layer-by-layer\footnote{A tree-structure is built-up from several
  layers (also called levels), with the top most level called the
  root, the bottom levels leaves and in between the nodes.} from top
to bottom. The construction of a single tree-level can be efficiently
vectorised which is required if one uses massively parallel
architectures.

\begin{figure}
\center
\includegraphics[width=\columnwidth]{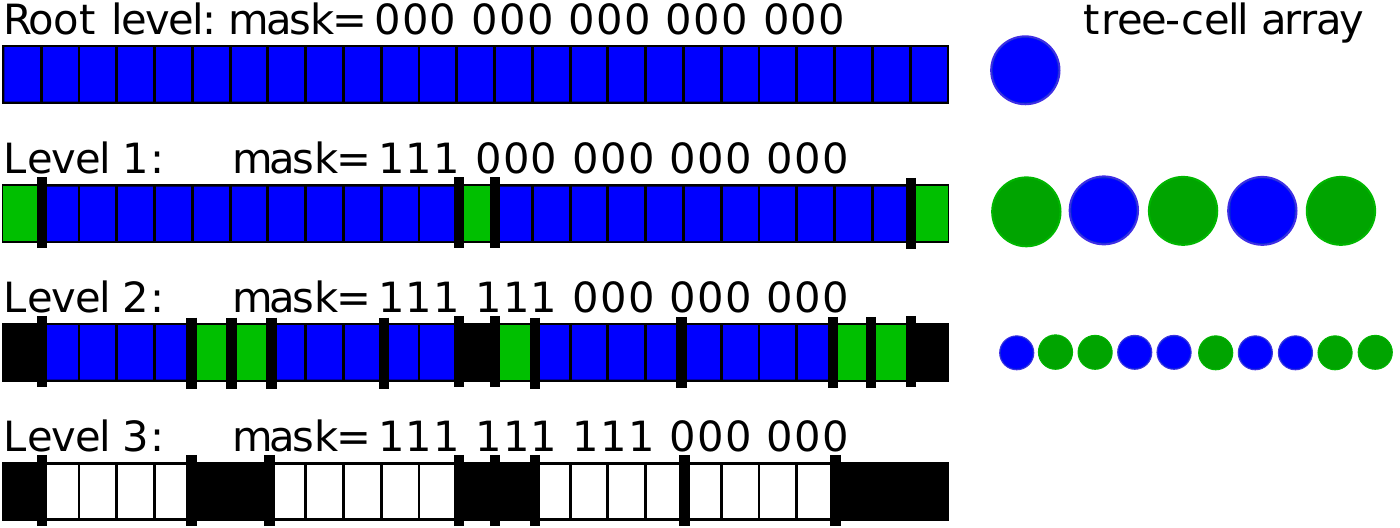}
\caption{Schematic representation of particle grouping into a tree cell. (Left panel) The particles
  Morton keys are masked$^6$ with the level mask. Next the particles with the same masked keys
  are grouped together into cells (indicated by a thick separator, such as in level 1). Cells with less than
  $N_{\rm leaf}$ particles are marked as leaves (green), and the corresponding particles are
  flagged (black boxes as in level 2 and 3) and not further used for the tree construction. 
  The other cells are called nodes (blue)
  and their particles are used constructing the next levels. The tree construction
  is complete as soon as all particles are assigned to leafs, or when the maximal depth of the tree
  is reached. (Right panel) The resulting array containing the created tree cells. }
\label{fig:treebuild}
\end{figure}

To vectorise the tree construction particles have to be mapped from a
3-dimensional spatial representation to a linear array while
preserving locality. This implies that particles that are nearby in 3D
space also have to be nearby in the 1D representation.  Space filling
curves, which trace through the three dimensional space of the data
enable such reordering of particles. The first use of space filling
curves in a tree-code was presented by Warren and Salmon (1993)
\cite{169640} to sort particles in a parallel tree-code for the
efficient distribution of particles over multiple systems.  This
sorting also improves the cache-efficiency of the tree-traverse since
most particles that are required during the interactions are stored
locally, which improves caching and reduces communication.  We adopt
the Morton space filling curve (also known as Z-order) \cite{Morton},
because of the existence of a one-to-one map between N-dimensional
coordinates and the corresponding 1D Morton-key. The Morton-keys give
a 1D representation of the original ND coordinate space and are
computed using bit-based operations (\ref{Sect:MKGen}). 
After the keys are
calculated the particles are sorted in increasing key order to achieve
a Z-ordered particle distribution in memory. The sorting is performed
using the radix-sort algorithm (see for our implementation details
\ref{Sect:Scan}), which we selected because of its better performance
compared to alternative sorting algorithms on the GPU
\cite{SatishSort, MerrillSorting}. After sorting the particles have to
be grouped into tree cells. In Fig.~\ref{fig:treebuild} (left panel), we
schematically demonstrate the procedure.  For a given level, we mask
the keys\footnote
{The masking is a bitwise operation that preserves the bits which are specified by the 
mask, for example masking ``1011b'' with ``1100b'' results in ``1000b''.} 
 of non-flagged particles (non-black elements of the array in
the figure), by assigning one particle per GPU thread. The thread
fetches the precomputed key, applies a mask (based on the
current tree level) and the result is the octree cell to which the 
particle should be assigned. Particles with identical masked keys 
are grouped together since they belong to the same cell. 
The grouping is implemented via the 
parallel compact algorithm (\ref{Sect:Scan}). We allow multiple
particles to be assigned to the same cell in order to reduce the size
of the tree-structure. The maximum number of particles that is
assigned to a cell is $N_{\rm leaf}$, which we set to $N_{\rm
  leaf}=16$. Cells containing up-to $N_{\rm leaf}$ particles are marked 
as leaves, otherwise they are marked as nodes. If a particle is assigned 
to a leaf the particle is flagged as complete (black elements of the array 
in the figure). The masking and grouping procedure is
repeated for every single level in serial until all particles are
assigned to leaves or that the maximal depth of the tree is reached,
whichever occurs first. When all particles are assigned to leaves 
all required tree cells have been created and are stored in a 
continuous array (right panel of Fig.~\ref{fig:treebuild}).

\begin{figure}
\center
\includegraphics[width=\columnwidth]{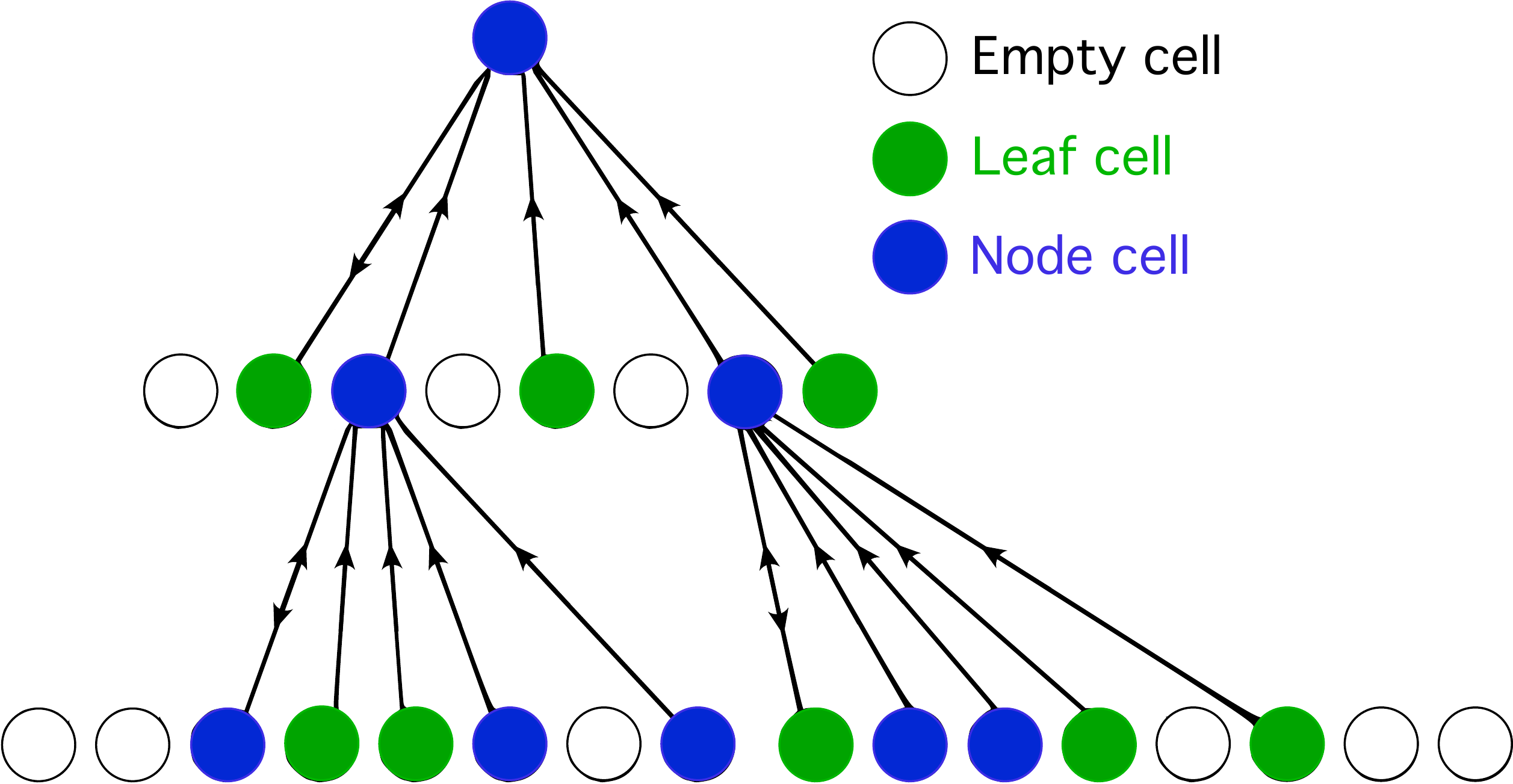}
\caption{Schematic illustration of the tree link process. Each cell of
  the tree, except the empty cells which are not stored, is assigned
  to a GPU thread. The thread locates the first child, if it exists,
  and the parent of the cell. The threads increment the child counter
  of the parent (indicated by the up arrows) and store the first child
  in the memory of the cell. This
  operation requires atomic read-modify-write operations because
  threads concurrently modify data at the same memory location.}
\label{fig:Link}
\end{figure}

However, to complete the tree-construction, the parent cells need to
be linked to their children. We use a separate function to connect the
parent and child cells with each other (linking the tree). This
function assigns a cell to a single thread which locates its parent
and the first child if the cell is not a leaf
(Fig.~\ref{fig:Link}). This is achieved by a binary search of the
appropriate Morton key in the array of tree-cells, which is already
sorted in increasing key order during the construction phase. The
thread increases the child counter of its parent cell and stores the
index of the first child. To reduce memory, we use a single integer to
store both the index to the first child and the number of children
(most significant 3 bits). If the cell is a leaf, we store the index
of the first particle instead of the index of the first child cell,
together with the number of particles in the leaf.  During these
operations many threads concurrently write data to the same memory
location.  To prevent race conditions, we apply atomic
read-modify-write instructions for modifying the data. At the end of
this step, the octree is complete and can be used to compute gravitational
attractions.

\subsection{Tree traverse}\label{Sect:TreeTrav}

To take advantage of the massively parallel nature of GPUs we use
breadth first,
instead of the more common depth first, traversal. Both breadth first
and depth first can be parallelised, but only the former can be
efficiently vectorised. To further vectorise the calculation we do not
traverse a single particle, but rather a group of particles. This
approach is known as Barnes' vectorisation of the tree-traverse
\cite{1990JCoPh..87..161B}.  The groups are based on the tree-cells to
take advantage of the particle locality as created during the
tree-construction. The number of particles in a group is $\leq N_{\rm
  crit}$ which is typically set to $64$ and the total number of 
groups is $N_{\rm groups}$. The groups are associated
with a GPU thread-block where the number of threads in a block is
$N_{\rm block}$ with $N_{\rm block} >= N_{\rm crit}$. Hereby we
assume that thousands of such blocks are able to run in parallel. This
assumption is valid for CUDA-enabled GPUs, as well as on AMD Radeon
GPUs via OpenCL.  If the code is executed with shared time-steps, all
particles are updated at the same time and subsequently all groups are
marked as active, for block time-steps this number varies between 1
and $N_{\rm groups}$. Each thread block executes the same
algorithm but on a different set of particles.

Each thread in a block reads particle data that belongs to the
corresponding group as well as group information which is required for
the tree traverse. If the number of particles assigned to a group is
smaller than $N_{\rm block}$ by a factor of two or more, we use
multiple threads (up to 4) per particle to further parallelise the
calculations. As soon as the particle and group data is read by the
threads we proceed with the tree-traverse.

\begin{figure}
\center
\includegraphics[width=\columnwidth]{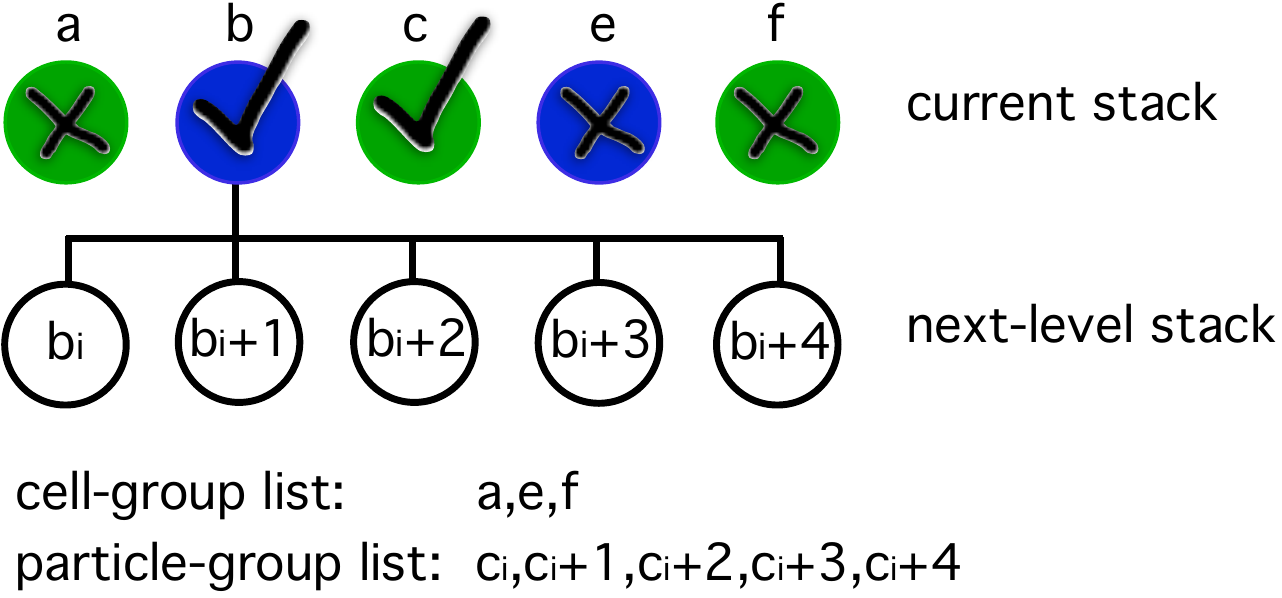}
\caption{Illustration of a single level tree-traverse. There are five
  cells in the current stack. Those cells which are marked with
  crosses terminate traverse, and therefore are added to the
  cell-group list for subsequent evaluation. Otherwise, if the cell is
  a node, its children are added to the next-level stack. Because
  children are contiguous in the tree-cell array, they are named as
  $b_i$, $b_i+1$, $\dots$, $b_i + 4$, where $b_i$ is the index of the
  first child of the node $b$. If the cell is a leaf, its particles
  are added to the particle-cell interaction list. Because particles
  in a leaf are also contiguous in memory, we only need to know the
  index of the first particle, $c_i$, and the number of particles in a
  leaf, which is 5 here.}
\label{fig:treewalk}
\end{figure}

On the CPU the tree-traverse algorithm is generally implemented using
recursion, but on the GPU this is not commonly supported and hard to
parallelise. Therefore we use a stack based breadth first
tree-traverse which allows parallelisation. Initially, cells from one
of the topmost levels of the tree are stored in the current-level
stack and the next-level stack is empty; in principle, this can be the
root level, but since it consists of one cell, the root node, only one
thread from $N_{\rm block}$ will be active. Taking a deeper level
prevents this and results in more parallelism. We loop over the cells
in the current-level stack with increments of $N_{\rm block}$. Within
the loop, the cells are distributed among the threads with no more
than one cell per thread. A thread reads the cell's properties and
tests whether or not to traverse the tree further down from this cell;
if so, and if the cell is a node the indexes of its children are added
to the next-level stack. If however, the cell is a leaf, the indexes
of constituent particles are stored in the particle-group interaction
list. Should the traverse be terminated then the cell itself is added
to cell-group interaction list (Fig.\ref{fig:treewalk}).

The cell-group interaction list is evaluated when the size of the list
exceeds $N_{\rm block}$. To achieve data parallelism each thread reads
properties of an interacting cell into fast low-latency on-chip memory
that can be shared between the threads, namely shared memory (CUDA) or
local memory (OpenCL).  Each thread then progresses over the data in
the on-chip memory and accumulates partial interactions on its
particle. At the end of this pass, the size of the interaction list is
decreased by $N_{\rm block}$ and a new pass is started until the size
of the list falls below $N_{\rm block}$. The particle-group
interaction list is evaluated in exactly the same way, except that
particle data is read into shared memory instead of cell data. This is
a standard data-sharing approach that has been used in a variety of
$N$-body codes, e.g. Nyland et al. \cite{Nyland_nbody}.

The tree-traverse loop is repeated until all the cells from the
current-level stack are processed.  If the next-level stack is empty,
the tree-traverse is complete, however if the next-level stack is
non-empty its data is copied to the current-level stack. The
next-level stack is cleaned and the current-level stack is
processed. When the tree-traverse is complete either the cell-group,
particle-group or both interaction lists may be non-empty. In such
case the elements in these lists are distributed among the threads and
evaluated in parallel. Finally if multiple threads per particle are
used an extra reduction step combines the results of these threads to
get the final interaction result.

\section{Gravitational Tree-code}\label{Sect:GravTree}

To demonstrate the feasibility and performance, we implement a version
of the gravitational Barnes-Hut tree-code \cite{1986Natur.324..446B}
which is solely based on the sparse octree methods presented in the
previous sections.  In contrast to other existing GPU codes
\cite{1654123, OctGravICCS10}, our implementation runs entirely on
GPUs. Apart from the previous described methods to construct and
traverse the tree-structure we implement time integration
(Section~\ref{Sect:PartInt}), time-step calculation and tree-cell
properties computation on the GPU (Section ~\ref{Sect:NodeProp}). The
cell opening method, which sets the accuracy and performance of the
tree-traverse, is described in Section~\ref{Sect:OpeningAngle}.

\subsection{Time Integration}\label{Sect:PartInt}

To move particles forward in time we apply the leapfrog
predictor-corrector scheme described by Hut et al. (1995)
\cite{1995ApJ...443L..93H}. Here the position and velocity are
predicted to the next simulation time using previously calculated
accelerations. Then the new accelerations are computed (tree-traverse)
and the velocities are corrected. This is done for all particles in
parallel or on a subset of particles in case of the block-time step
regime.  For a cluster of $\apgt 10^5$ particles, the time required
for one prediction-correction step is less than 1\% of the total
execution time and therefore negligible.

\subsection{Tree-cell properties}\label{Sect:NodeProp}

\begin{figure}
\center
\includegraphics[width=\columnwidth]{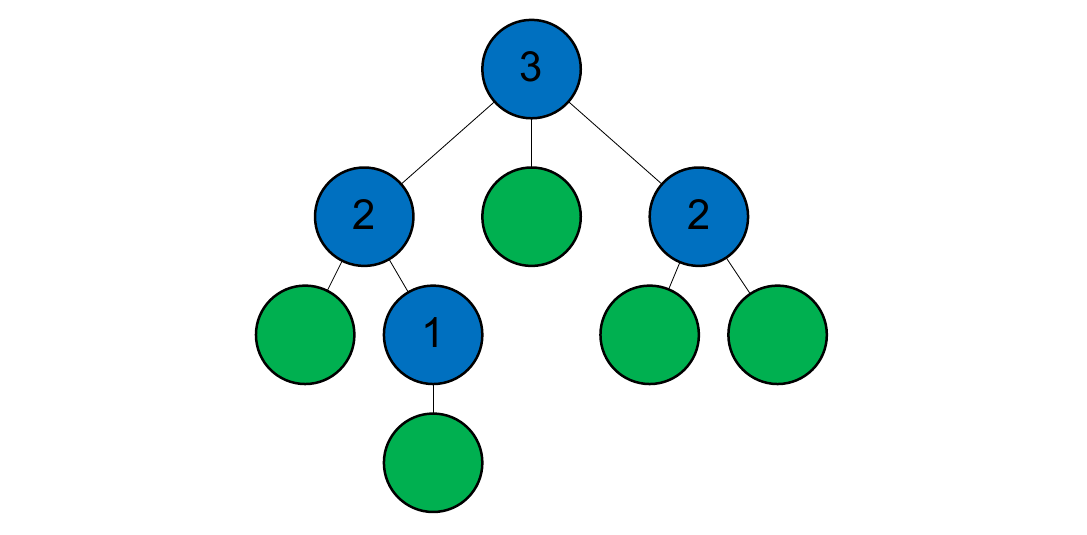}
\caption{Illustration of the computation of the multipole moments. First the properties
of the leaves are calculated (green circles). Then the properties of the nodes are
calculated level-by-level from bottom to top. This is indicated by the numbers in the nodes, 
first we compute the properties of the node with number 1, followed by the nodes with number 
2 and finally the root node.}
\label{fig:NodeProps}
\end{figure}

Tree-cell properties are a summarized representation of the underlying
particle distribution. The multipole moments are used to compute the
forces between tree-cells and the particles that traverse the tree. In
this implementation of the Barnes-Hut tree-code we use only monopole
and quadrupole moments \cite{1993ApJ...414..200M}. Multipole moments
are computed from particle positions and need to be recomputed at each
time-step; any slowdown in their computation may substantially
influence the execution time.  To parallelise this process we
initially compute the multipole moments of each leaf in parallel. We
subsequently traverse the tree from bottom to top. At each level the
multipole moments of the nodes are computed in parallel using the
moments of the cells one level below (Fig.~\ref{fig:NodeProps}). The
number of GPU threads used per level is equal to the number of nodes
at that level.  These computations are performed in double precision
since our tests indicated that the NVIDIA compiler aggressively
optimises single precision arithmetic operations, which results in an
error of at most 1\% in the multipole moments. Double precision
arithmetic solved this problem and since the functions are
memory-bound\footnote{On GPUs we distinguish two kind of performance
  limitations, memory-bound and compute-bound. In the former the
  performance is limited by the memory speed and memory bandwidth, in
  the later the performance is limited by the computation speed.}  the
overhead is less than a factor 2.  As final step the double precision
values are converted back to single precision to be used during the
tree-traverse.

\subsection{Cell opening criterion}\label{Sect:OpeningAngle}

In a gravitational tree-code the multipole acceptance criterion (MAC)
decides whether to use the multipole moments of a cell or to further
traverse the tree. This influences the precision and execution time of 
the tree-code. The further the tree is traversed the more accurate the
computed acceleration will be. However, traversing the tree further results in
a higher execution time since the number of gravitational interactions
increases. Therefore the choice of MAC is important, since it tries
to determine, giving a set of parameters, when the distance between
a particle and a tree cell is large enough that the resulting force
approximation error is small enough to be negligible.
  The MAC used in this work is a combination of the
method introduced by Barnes (1994) \cite{Barnes1994} and the method
used for tree-traversal on vector machines
\cite{1990JCoPh..87..161B}. This gives the following acceptance
criterion,
\begin{equation}
\label{Eq:Opening}
d > {l\over{\theta}} + \delta
\end{equation}
where $d$ is the smallest distance between a group and the center of
mass of the cell, $l$ is the size of the cell, $\theta$ is a dimensionless
parameter that controls the accuracy and $\delta$ is the distance 
between the cell's geometrical center and the center of mass. If
$d$ is larger than the right side of the equation the distance is 
large enough to use the multipole moment instead of traversing
to the child cells.

\begin{figure}
\center
\includegraphics[width=\columnwidth]{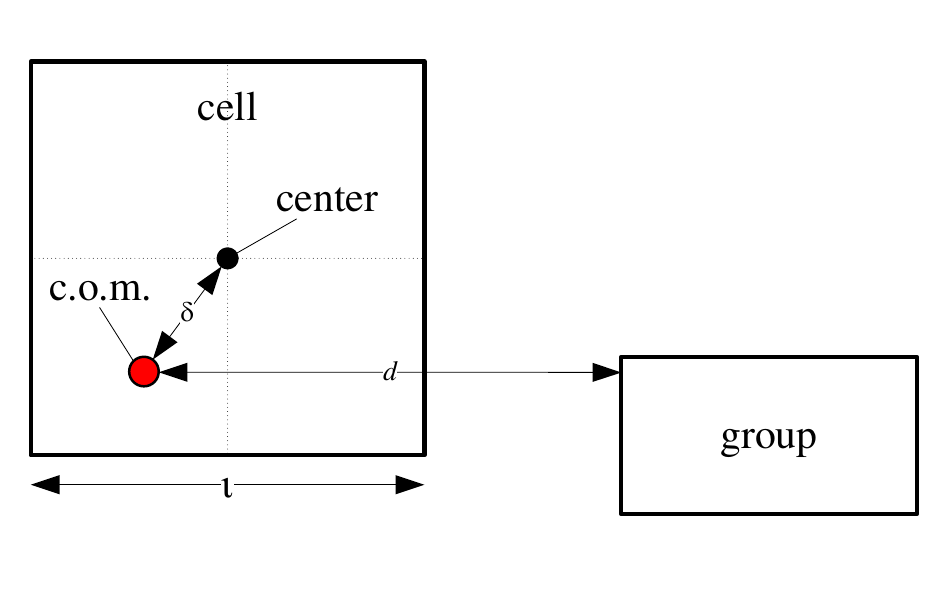}
\caption{Schematic overview of the multipole acceptance criterion.  We
  determine if the cell has to be opened (continue tree-traversal)
  for the whole group and
  therefore take the minimal distance $d$ between the group and the
  center of mass of the node.  The cell size and the distance between
  the center of mass and the geometrical center are indicated by $l$
  and $\delta$ respectively. The parameters are used in Eq.~\ref{Eq:Opening} 
  to determine if the cell has to be opened. }
\label{fig:OpeningCriteria}
\end{figure}

Fig.~\ref{fig:OpeningCriteria} gives
an overview of this method. We also implemented the minimal distance
MAC \cite{1994JCoPh.111..136S}, which results in an acceleration
error that is between 10\%~and~50\% smaller for the same $\theta$ than
the MAC used here.  The computation time, however, is almost a factor
3 higher since more cells are accepted (opened).

The accuracy of the tree-traverse is controlled by the parameter $\theta$.
Larger values of $\theta$ causes fewer cells to be opened and
consequently results in a shallower tree-traverse and a faster evaluation of
the underlying simulation. Smaller values of $\theta$ have the exact
opposite effect, but result in a more accurate integration.  In the
hypothetical case that all the tree cells are opened ($\theta
\rightarrow 0$) the tree-code turns in an inefficient direct $N$-body
code. In Section~\ref{Sect:PerfRes} we adopt $\theta=0.5$ and
$\theta=0.75$ to show the dependence of the execution time on the
opening angle. In Section~\ref{Sect:AccRes} we vary $\theta$ between
0.2 and 0.8 to show the dependence of the acceleration error on
$\theta$.

\begin{table}[!t]
\caption{Used hardware. The Xeon is the CPU in the host system, the 
other five devices are GPUs.}
\label{Tab:GPUProps}
\begin{center}
    \begin{tabular}{ | l | c | c | c | c | c | c |}
    \hline
    Hardware model      & Xeon E5620  & 8800 GTS512 & C1060 & GTX285 & C2050 & GTX480    \\ \hline
    Architecture        & Gulftown  & G92 & GT200 & GT200 & GF100 & GF100    \\ \hline \hline
    \#  Cores           &  4     & 128  & 240  & 240  & 448  & 480\\ \hline
    Core (Mhz)          &  2400   & 1625  & 1296 & 1476 & 1150 & 1401 \\ \hline
    Memory (Mhz)        &  1066   & 1000  & 800  & 1243 & 1550 & 1848\\ \hline
    Interface (bit)     &  192    & 256   & 512  & 512  & 384 & 384 \\ \hline
    Bandwidth (GBs)     & 25.6    & 64    & 102  & 159  & 148 & 177.4 \\ \hline
    Peak (GFLOPs)\footnotemark[2] & 76.8   & 624   &  933 & 1063 & 1030 & 1345 \\ \hline
    Memory size (GB)    & 16    & 0.5   &  4 & 1 & 2.5 & 1.5 \\ \hline
    \end{tabular}
\end{center}
\footnotemark[1]{All calculations in this work are done in single precision arithmetic.} \\ 
\footnotemark[2]{The peak performance is calculated as follows: \\
Gulftown:\space\space\space\space\space\space     \#Cores $\times$ Core speed $\times$ 8 (SSE flops/cycle)  \\
G92 \& GT200: \#Cores $\times$ Core speed $\times$ 3 (flops/cycle)  \\
GF100: \space\space\space\space\space\space\hspace{1mm}     \#Cores $\times$ Core speed $\times$ 2 (flops/cycle) }

\end{table}

\section{Performance and Accuracy}\label{Sect:Results}

In this section we compare the performance of our implementation of
the gravitational $N$-body code ({\tt Bonsai}) with
CPU implementations of comparable algorithms.  Furthermore, we use a
statistical test to compare the accuracy of {\tt Bonsai} with a direct
summation code. As final test {\tt Bonsai} is compared with a direct
$N$-body code and a set of $N$-body tree-codes using a production type
galaxy merger simulation.

Even though there are quite a number of tree-code implementations each
has its own specific details and it is therefore difficult to give a
one-to-one comparison with other tree-codes. The implementations
closest to this work are the parallel CPU tree-code of John Dubinski
(1996) \cite{1996NewA....1..133D} ({\tt Partree}) and the GPU
accelerated tree-code {\tt Octgrav} \cite{OctGravICCS10}. Other often
used tree-codes either have a different MAC or lack quadrupole
corrections.  The default version of {\tt Octgrav} has a different MAC
than {\tt Bonsai}, but for the galaxy merger simulation a version of
{\tt Octgrav} is used that employs the same method as {\tt Bonsai}
(Section~\ref{Sect:OpeningAngle}).  We use {\tt phiGRAPE}
\cite{2007NewA...12..357H} in combination with the {\tt Sapporo}
\cite{Gaburov2009630} direct $N$-body GPU library for the comparison
with direct $N$-body simulations. Although here used as standalone 
codes, most of them are part of the AMUSE 
framework\cite{2009NewA...14..369P}, as will be a
future version of {\tt Bonsai} which would make the comparison
trivial to execute. 

The hardware used to run the tests is presented in
Table~\ref{Tab:GPUProps}. For the CPU calculations we used an Intel
Xeon E5620 CPU which has 4 physical cores.  For GPUs, we used 1 GPU
with the G92 architecture (GeForce 8800GTS512), 2 GPUs with the GT200
architecture (GeForce 285GTX and Tesla C1060), and 2 GPUs with the
GF100 architecture (GTX480 and Tesla C2050). All these GPUs are
produced by NVIDIA.  The Tesla C2050 GPU is marketed as a professional
High Performance Computing card and has the option to enable
error-correcting code memory (ECC). With ECC enabled extra checks on
the data are conducted to prevent the use of corrupted data in
computations, but this has a measurable impact on the
performance. Therefore, the tests on the C2050 are executed twice,
once with and once without ECC enabled to measure the impact of ECC.

All calculations are conducted in single precision arithmetic except
for the computation of the monopole and quadrupole moments in {\tt
  Bonsai} and the force calculation during the acceleration test in
{\tt phiGRAPE} for which we use double precision arithmetic.

\subsection{Performance}\label{Sect:PerfRes}

\begin{figure}
\center
\includegraphics[width=\columnwidth]{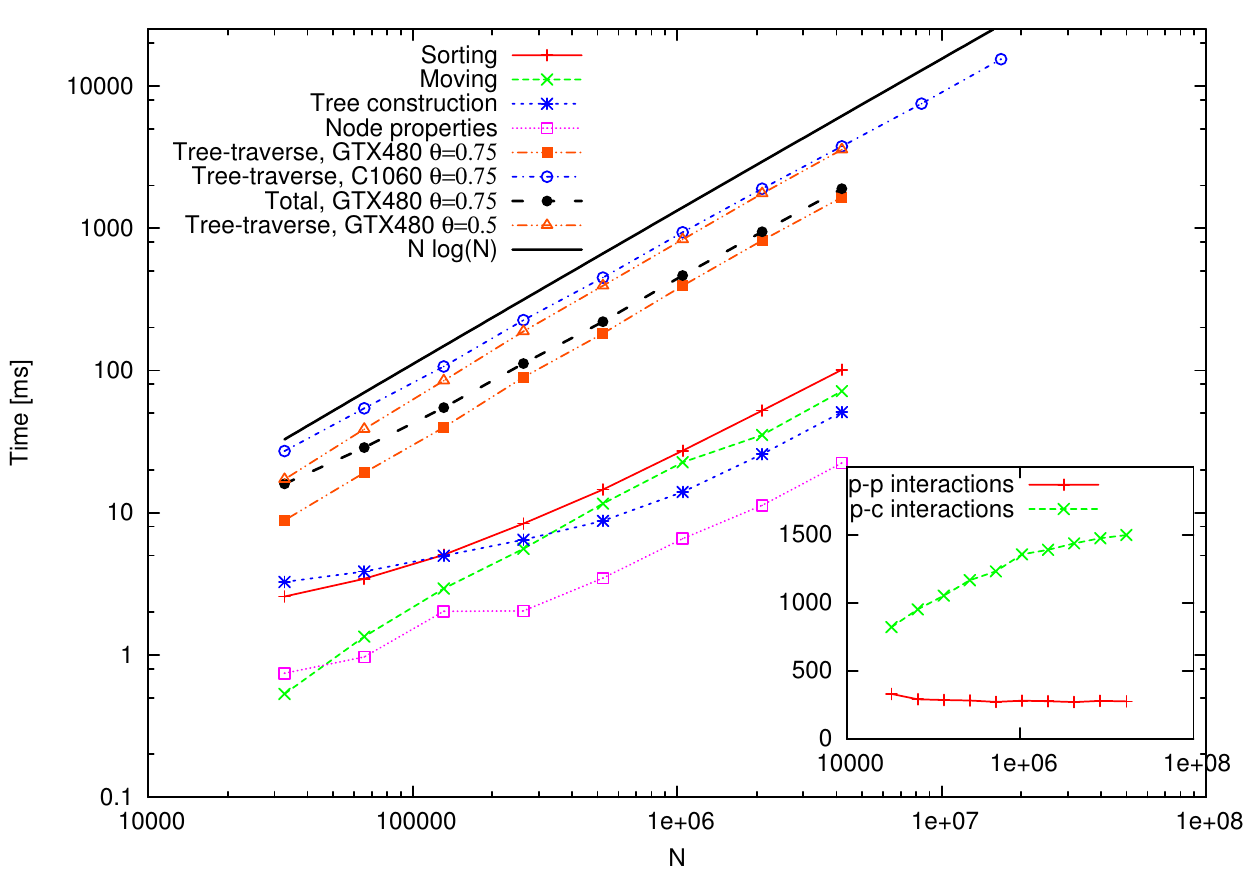}
\caption{The wall-clock time spent by various parts of the
  program versus the number of particles $N$. We used Plummer models as initial conditions
  \cite{1915MNRAS..76..107P} and varied the number of particles over two orders of
  magnitude. The solid black line, which is offset arbitrarily, shows the theoretical ${\cal O}(N\log N)$ scaling \cite{1986Natur.324..446B}. 
  The asymptotic complexity of the tree-construction approaches ${\cal O}(N)$, 
  because all the constituent primitives share the same complexity.   
  The tree-construction timing comes from the GTX480.
  To show that the linear scaling continues we added timing data for the C1060, 
  which allows usage of larger data sets. For the GTX480 we included the results of 
  the tree-traverse with $\theta$ = 0.5 and the results of the tree-traverse 
  with $\theta$ = 0.75.
  The inset shows the average number of particle-particle
  and particle-cell interactions for each simulation where $\theta$ = 0.75.}
\label{Image:Scaling}
\end{figure}

To measure the performance of the implemented algorithms we execute 
simulations using Plummer \cite{1915MNRAS..76..107P} spheres
with $N=2^{15}$ (~32k) up to $N=2^{24}$ (~16M)
particles (up to $N=2^{22}$ (~4M) for the GTX480, because of memory
limitations). For the most time critical parts of the algorithm we 
measure the wall-clock time. For the tree-construction we distinguish
three parts, namely sorting of the
keys (sorting), reordering of particle properties based on the sorted
keys (moving) and construction and linking of tree-cells
(tree-construction). Furthermore, are timings presented for the
multipole computation and tree-traverse. The results are presented in 
Fig.~\ref{Image:Scaling}.
The wall-clock time spend in the sorting, moving,
tree-construction and multipole computation algorithms scales linearly
with $N$ for $N \gtrsim 10^6$.  For smaller $N$, however, the scaling
is sub-linear, because the parallel scan algorithms require more than
$10^5$ particles to saturate the GPU.  
The inset of Fig.~\ref{Image:Scaling} shows that the average number
of particle-cell interactions doubles between $N \gtrsim $~32k and 
$N \lesssim $~1M and keeps gradually increasing for $N \gtrsim $~1M.
Finally, more than 90\% with
$\theta=0.75$ and 95\% with $\theta=0.5$ of the wall-clock time is
spent on tree-traversal. This allows for block time-step execution
where the tree-traverse time is reduced by a factor 
$N_{\rm groups}/N_{\rm active}$, where $N_{\rm active}$ is the 
number of groups with particles that have to be updated.

\begin{figure}
\center \includegraphics[width=\columnwidth]{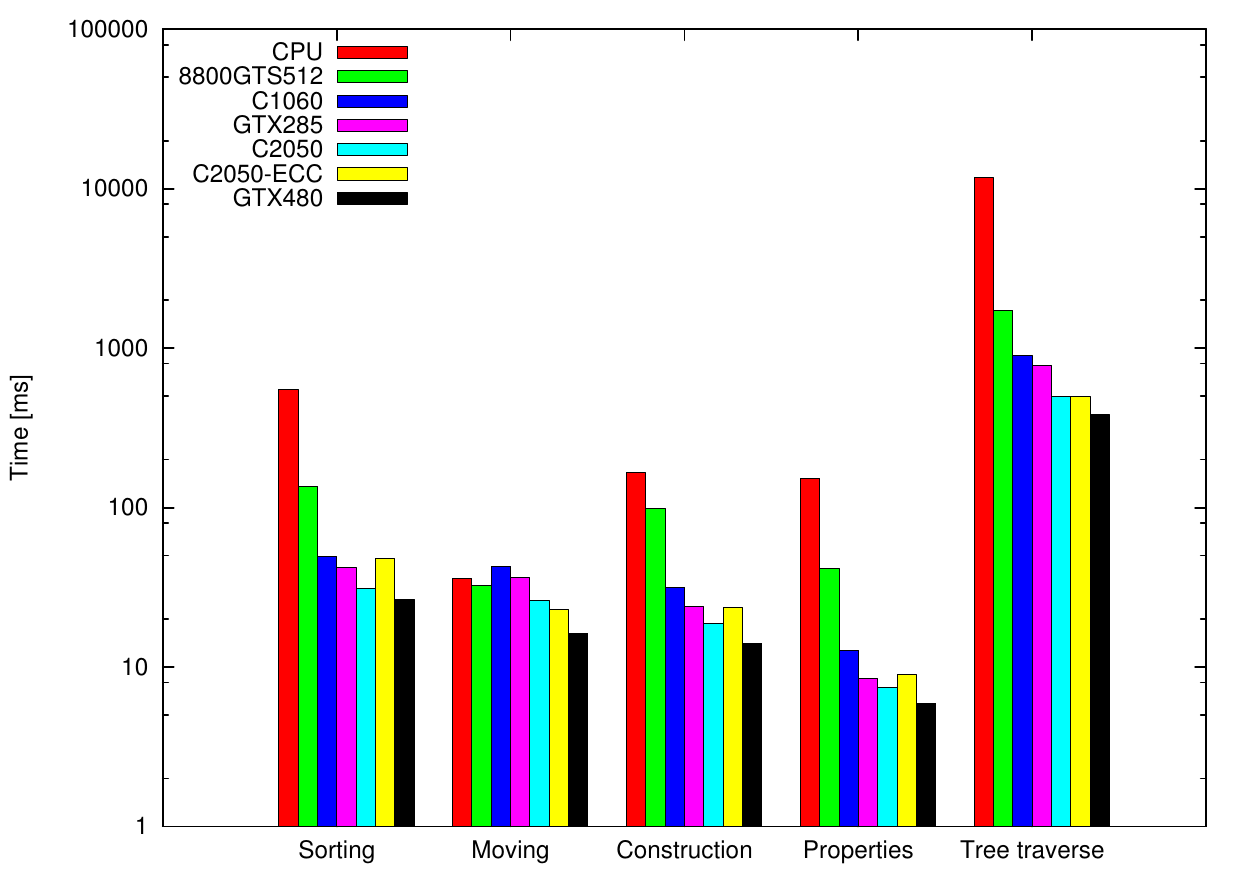}
\caption{Wall-clock time spent by the CPU and different generations of
  GPUs on various primitive algorithms. The bars show the time spent
  on the five selected sections of code on the CPU and 5 different
  GPUs (spread over 3 generations). The results indicate that the code
  outperforms the CPU on all fronts, and scales in a predictable
  manner between different GPUs. The {\tt C2050-ECC} bars indicate the
  runs on the C2050 with ECC enabled, the {\tt C2050} bars indicate
  the runs with ECC disabled.  Note that the y-axis is in log scale.
 (Timings using a $2^{20}$ million body Plummer sphere with $\theta=0.75$)}
\label{Image:GenScaling}
\end{figure}

In Fig.~\ref{Image:GenScaling} we compare the performance of the
tree-algorithms between the three generations of GPUs as well as
against tuned CPU implementations\footnote{ The tree-construction
  method is similar to \cite{169640}, and was implemented by Keigo
  Nitadori with OpenMP and SSE support.  The tree-traverse is,
  however, from the CPU version of the MPI-parallel tree-code by John
  Dubinski \cite{1996NewA....1..133D}. It has monopole and quadrupole
  moments and uses the same multipole acceptance criterion as our
  code. We ran this code on the Xeon E5620 CPU using 4 parallel
  processes where each process uses one of the 4 available physical
  cores.}.  For all algorithms the CPU is between a factor of 2 (data
reordering) to almost a factor 30 (tree-traverse) slower than the
fastest GPU (GTX480).  Comparing the results of the different GPUs we
see that the GTS512 is slowest in all algorithms except for the data
moving phase, in which the C1060 is the slowest. This is surprising
since the C1060 has more on-device bandwidth, but the lower memory
clock-speed appears to have more influence than the total bandwidth.
Overall the GF100 generation of GPUs have the best performance. In
particular, during the tree-traverse part, they are almost a factor 2
faster than the GT200 series.  This is more than their theoretical
peak performance ratios, which are 1.1 and 1.25 for C1060 vs. C2050
and GTX285 vs. GTX480 respectively. In contrast, the GTX285 executes
the tree-traverse faster than the C1060 by a factor of 1.1  which is exactly
the peak performance ratio between these GPUs.  We believe that the
difference between the GT200 and GF100 GPUs is mainly caused by the
lack of L1 and L2 caches on GT200 GPUs that are present on GF100 GPUs.
In the latter, non-coalesced memory accesses are cached, which occur
frequently during the tree-traverse, this reduces the need to request
data from the relatively slow global memory.  This is supported by
auxiliary tests where the texture cache on the GT200 GPUs is used to
cache non-coalesced memory reads, which resulted in a reduction of the
tree-traverse execution time between 20 and 30\%.  Comparing the C2050
results with ECC-memory to those without ECC-memory we notice a
performance impact on memory-bound functions that can be as high as
50\% (sorting), while the impact on the compute-bound tree-traverse is
negligible, because the time to perform the ECC is hidden behind
computations.  Overall we find that the implementation scales very
well over the different GPU generations and makes optimal use of the
newly introduced features of the GF100 architecture. The performance
of the tree-traverse with $\theta=0.75$ is 2.1M particles/s and 2.8M
particles/s on the C2050 and GTX480 respectively for N = 1M.

\subsection{Accuracy}\label{Sect:AccRes}

To measure the accuracy of the tree-code we use two tests.  In the
first, the accelerations due to the tree-code are compared with
accelerations computed by direct summation. In the second test, we
compared the performance and accuracy of three tree-codes and a direct
summation code using a galaxy merger simulation.

\subsubsection{Acceleration}

To quantify the error in the accelerations between {\tt phiGRAPE} and
{\tt Bonsai} we calculate
\begin{equation} \label{Eq:AccErr}
 \Delta a/a = |{\bf a_{\rm tree} - a_{\rm direct}}|/|{\bf a_{\rm direct}}|,
\end{equation}
where ${\bf a}_{\rm tree}$ and ${\bf a}_{\rm direct}$ are
accelerations obtained by tree and direct summation respectively. The
direct summation results are computed with a double precision version
of {\tt Sapporo}, while for tree summation single precision is
used. For both methods the softening is set to zero and a GTX480 GPU
is used as computation device.

\begin{figure}
\center \includegraphics[width=\columnwidth]{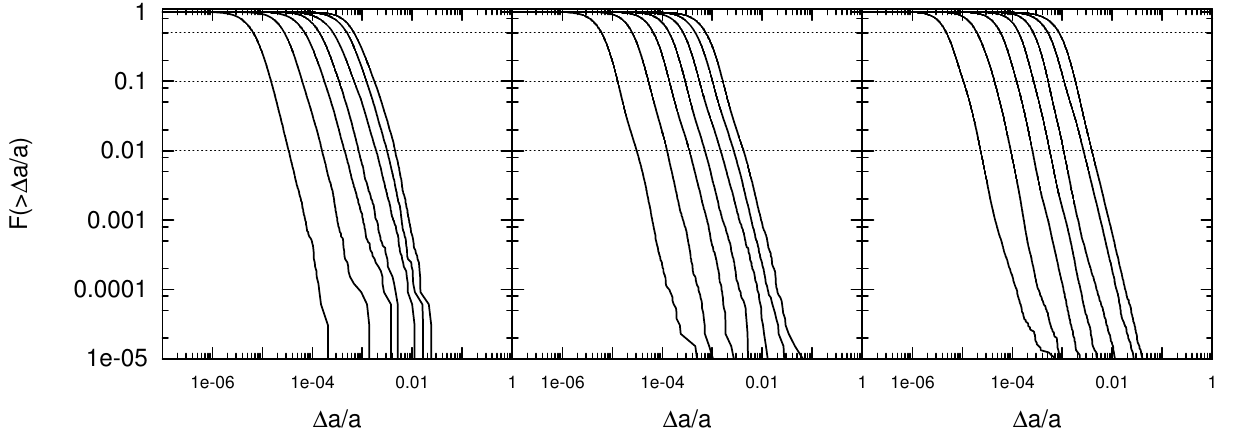}
\caption{ Each panel displays a fraction of particles, $F(> \Delta a/a)$,
  having a
  relative acceleration error, $\Delta a/a$, (vertical axis) greater than a specified
  value (horizontal axis). In each panel the solid lines show the
  errors for various opening angles, from left to right $\theta=0.2$,
  0.3, 0.4, 0.5, 0.6, 0.7 and 0.8.  The panels show, from left to
  right, simulations with $N=32768$, $N=131072$ and $N=1048576$
  particles.  The dotted horizontal lines indicate 50\%, 10\% and 1\%
  of the error distribution.  }
\label{Image:AccAccel}
\end{figure}

In Fig. \ref{Image:AccAccel} the error distribution for different
particle numbers and opening angles is shown. Each panel shows the
fraction of particles (vertical-axis) having a relative acceleration
error larger than a given value (horizontal-axis).  The three
horizontal dotted lines show the 50th, 10th and 1st percentile of the
cumulative distribution (top to bottom).  The results indicate that
the acceleration error is slightly smaller (less than an order of
magnitude) than {\tt Octgrav} and comparable to CPU tree-codes with
quadrupole corrections \cite{2002JCoPh.179...27D, 2001NewA....6...79S,
  2001PhDT........21S}.  In {\tt Octgrav} a different MAC is used than
in {\tt Bonsai} which explains the better accuracy results of {\tt
  Bonsai}.

\begin{figure}
\center \includegraphics[width=\columnwidth]{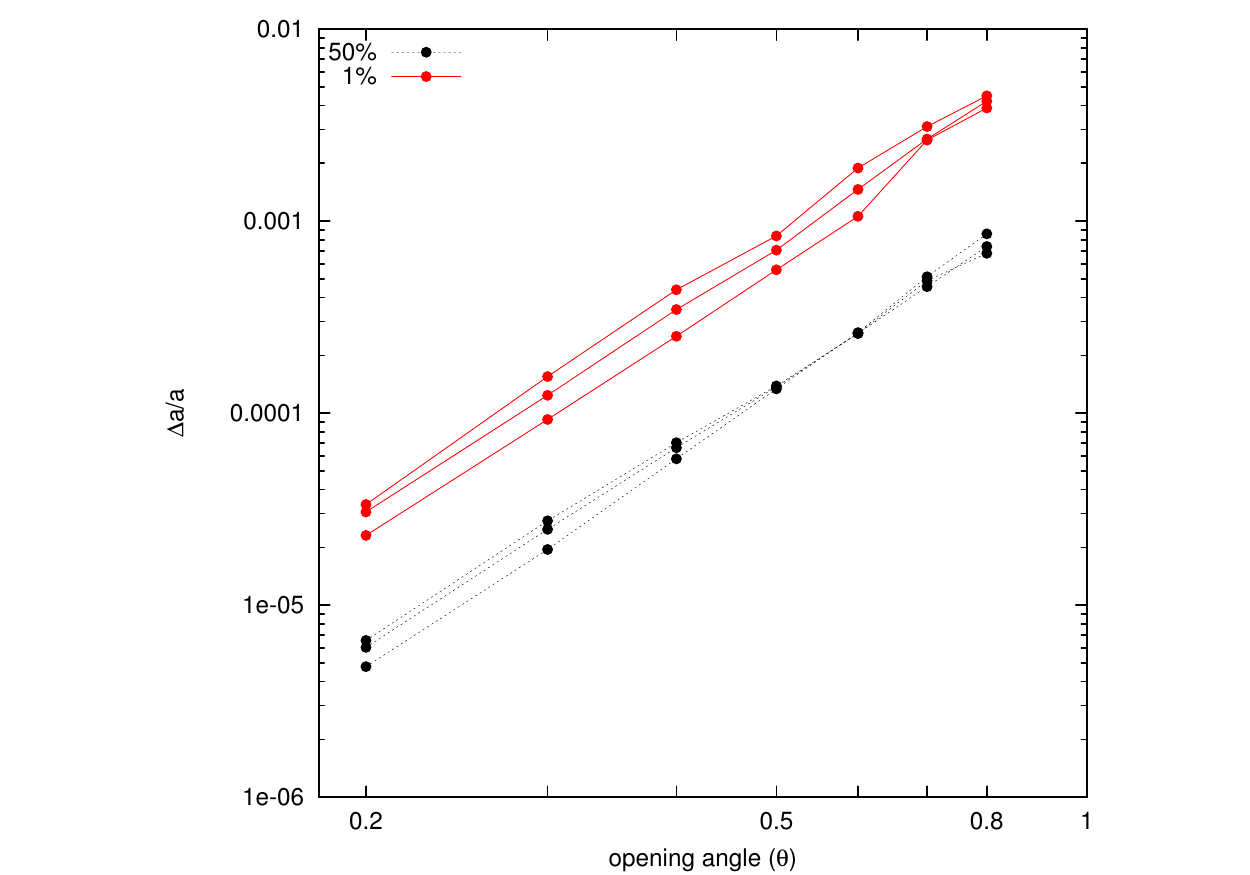}
\caption{ The median and the first percentile of the relative
  acceleration error distribution as a function of the opening angle
  and the number of particles. We show the lines for $N=32768$ (bottom
  striped and solid line) $N=131072$ (middle striped and dotted line)
  and $N=1048576$ (top striped and dotted line).  }
\label{Image:AccAccel2}
\end{figure}

The dependence of the acceleration error on $\theta$ and the number of
particles is shown in Fig.~\ref{Image:AccAccel2}. Here the median and
first percentile of the relative acceleration error distributions of
Fig.~\ref{Image:AccAccel} are plotted as a function of $\theta$. The
figure shows that the relative acceleration error is nearly
independent of $N$, which is a major improvement compared to {\tt
  Octgrav} where the relative acceleration error clearly depends on
$N$ (Figure~5 in \cite{OctGravICCS10}). The results are consistent with
those of {\tt Partree} \cite{1996NewA....1..133D} which uses the same MAC.


\subsubsection{Galaxy merger}

A realistic comparison between the different $N$-body codes, instead
of statistical tests only, is performed by executing a galaxy merger
simulation. The merger consists of two galaxies, each with $10^5$ dark
matter particles, $2\times10^4$ star particles and one super massive
black hole (for a total of 240.002 bodies). 
The galaxies have a 1:3 mass ratio and the pericenter is
10kpc. The merger is simulated with
{\tt Bonsai, Octgrav, Partree} and {\tt phiGRAPE}. The used hardware
for {\tt Bonsai} and {\tt Octgrav} was 1 GTX480, {\tt Partree} used 4
cores of the Intel Xeon E5620 and {\tt phiGRAPE} used 4 GTX480 GPUs.
With each tree-code we ran two simulations, one with $\theta=0.5$ and
one with $\theta=0.75$.  The end-time of the simulations is $T=1000$ 
with a shared time-step of $1\over64$ (resulting in 64000 steps) and
a gravitational softening of $\epsilon=0.1$. 
 For {\tt phiGRAPE} the default time-step settings were
used, with the maximum time-step set to $1\over16$ and softening set
to $\epsilon=0.1$. The settings are summarised in the first four
columns of Table~\ref{Tab:MergerProps}.

\begin{figure}
\center \includegraphics[width=\columnwidth]{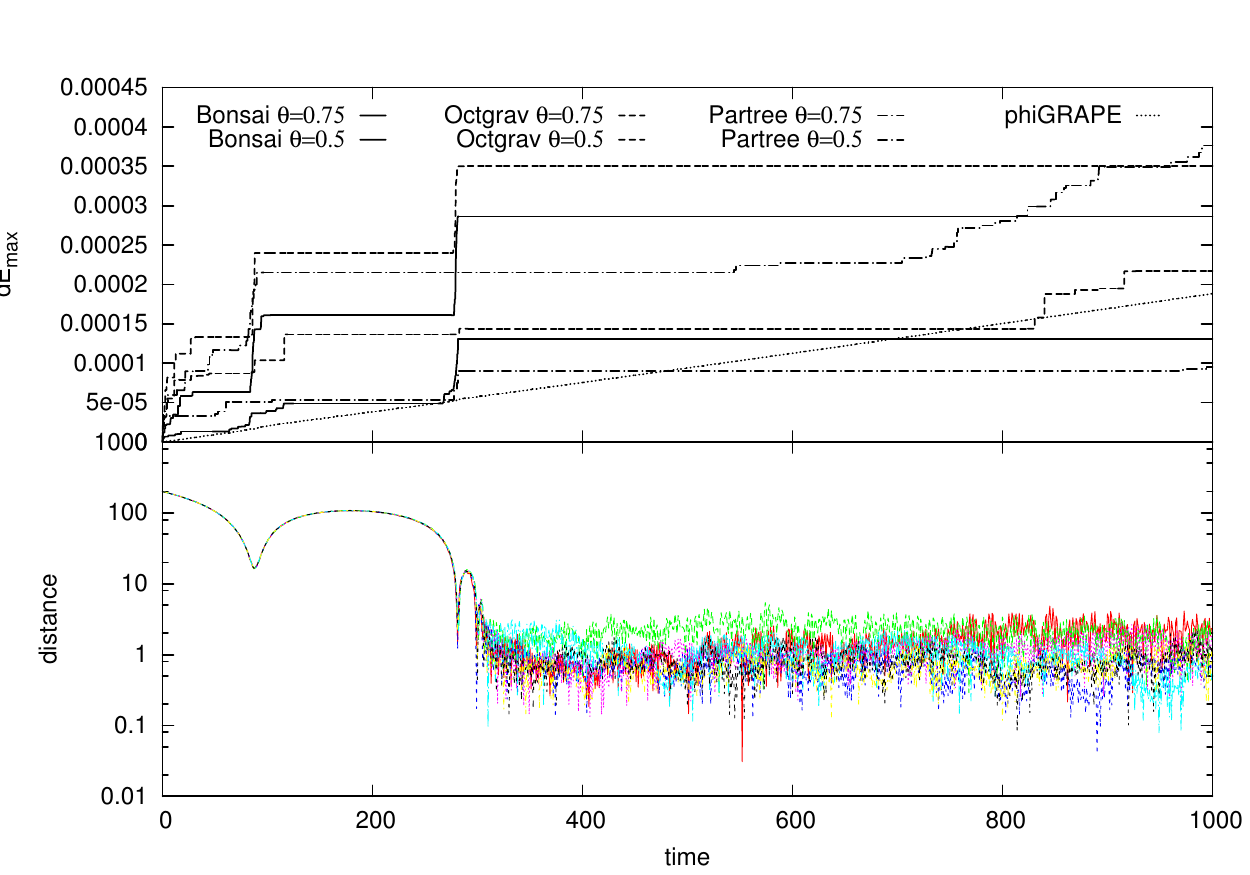}
\caption{
The top panel shows the maximal relative energy error over the 
course of the simulation. The solid lines show the results of {\tt Bonsai}, the striped lines
the results of {\tt Octgrav} and the striped-dotted lines show the results of {\tt Partree}. In
all cases the top lines show the result of $\theta=0.75$ and the bottom lines the result of 
$\theta=0.5$. Finally the dotted line shows the result of {\tt phiGRAPE}.
The bottom panel shows the distance between the two supermassive black-holes.
The lines themselves have no labels since up to $t=300$ the lines follow the same path 
and after $t=300$ the motion becomes chaotic and incomparable. 
The distance and time are in $N$-body units.
}
\label{Image:MergComp2}
\end{figure}

We compared the density, cumulative mass and circular velocity 
profiles of the  merger product as produced by the different simulation codes, 
but apart from slight differences caused by small number statistics the 
profiles are identical. As final comparison we recorded the distance between 
the two black holes over the course of the simulation. The results of which 
are shown in the bottom panel of Fig.~\ref{Image:MergComp2}, the results are
indistinguishable up to the third pericenter passage at $t=300$
after which the results, because of numerical differences, become incomparable.
Apart from the simulation results we also compare
the energy conservation. This is done by computing the relative energy
error ($dE$, Eq.~\ref{Eq:de}) and the maximal relative energy error
($dE_{max}$).

\begin{table}[!t]
\caption{Settings and results of the galaxy merger. The first two columns indicate the 
software and hardware used, the third the time-step (dt) and the fourth 
the opening angle ($\theta$). 
The last three columns present the results, energy error at the $time=1000$
(fifth column), maximum energy error during the simulation 
(sixth column) and the total execution time (seventh column).}
\label{Tab:MergerProps}
\begin{center}
    \begin{tabular}{ | l | c | c | c | c | c | l |}

    \hline
    Simulation          & Hardware   & dt             & $\theta$ & \multicolumn{1}{c|}{$dE_{end}[\times10^{-4}]$} & \multicolumn{1}{c|}{$dE_{max}[\times10^{-4}]$} & time[s]    \\ \hline    
    phiGRAPE            & 4x GTX480  & block          & -        & $-1.9$  & $1.8$   & 62068\\ \hline
    Bonsai  run1        & 1x GTX480  & $\frac{1}{64}$ & $0.75$   & $ 0.21$  & $2.8$ & 7102 \\ \hline
    Bonsai  run2        & 1x GTX480  & $\frac{1}{64}$ & $0.50$    & $ 0.44$  & $1.3$ & 12687\\ \hline
    Octgrav run1        & 1x GTX480  & $\frac{1}{64}$ & $0.75$   & $-1.1$  & $3.5$ & 11651 \\ \hline
    Octgrav run2        & 1x GTX480  & $\frac{1}{64}$ & $0.50$    & $-1.1$  & $2.2$ & 15958 \\ \hline
    Patree  run1        & Xeon E5620 & $\frac{1}{64}$ & $0.75$   & $-3.5$  & $3.8$ & 118424 \\ \hline    
    Patree  run2        & Xeon E5620 & $\frac{1}{64}$ & $0.50$    & $0.87$   & $0.96$ & 303504 \\ \hline   
    \end{tabular}
\end{center}
\end{table}

\begin{equation} \label{Eq:de}
dE = {E_0-E_t\over{E_0}}
\end{equation}

Here $E_0$ is the total energy (potential energy + kinetic energy) at
the start of the simulation and $E_t$ is the total energy at time $t$.
The time is in $N$-body units.

The maximal relative energy error is presented in the top panel of
Fig.~\ref{Image:MergComp2}, the tree-code simulations with
$\theta=0.75$ give the highest $dE_{max}$ which occurs for both
$\theta=0.75$ and $\theta=0.5$ during the second pericenter passage at
$t\approx280$. For $\theta=0.5$, the $dE_{max}$ is roughly a factor 2
smaller than for $\theta=0.75$.  For {\tt phiGRAPE} $dE_{max}$ shows a
drift and does not stay constant after the second pericenter
passage. The drift in the energy error is caused by the formation of
binaries, for which {\tt phiGRAPE} has no special treatment, resulting
in the observed drift instead of a random walk.

\begin{figure}
\center \includegraphics[width=\columnwidth]{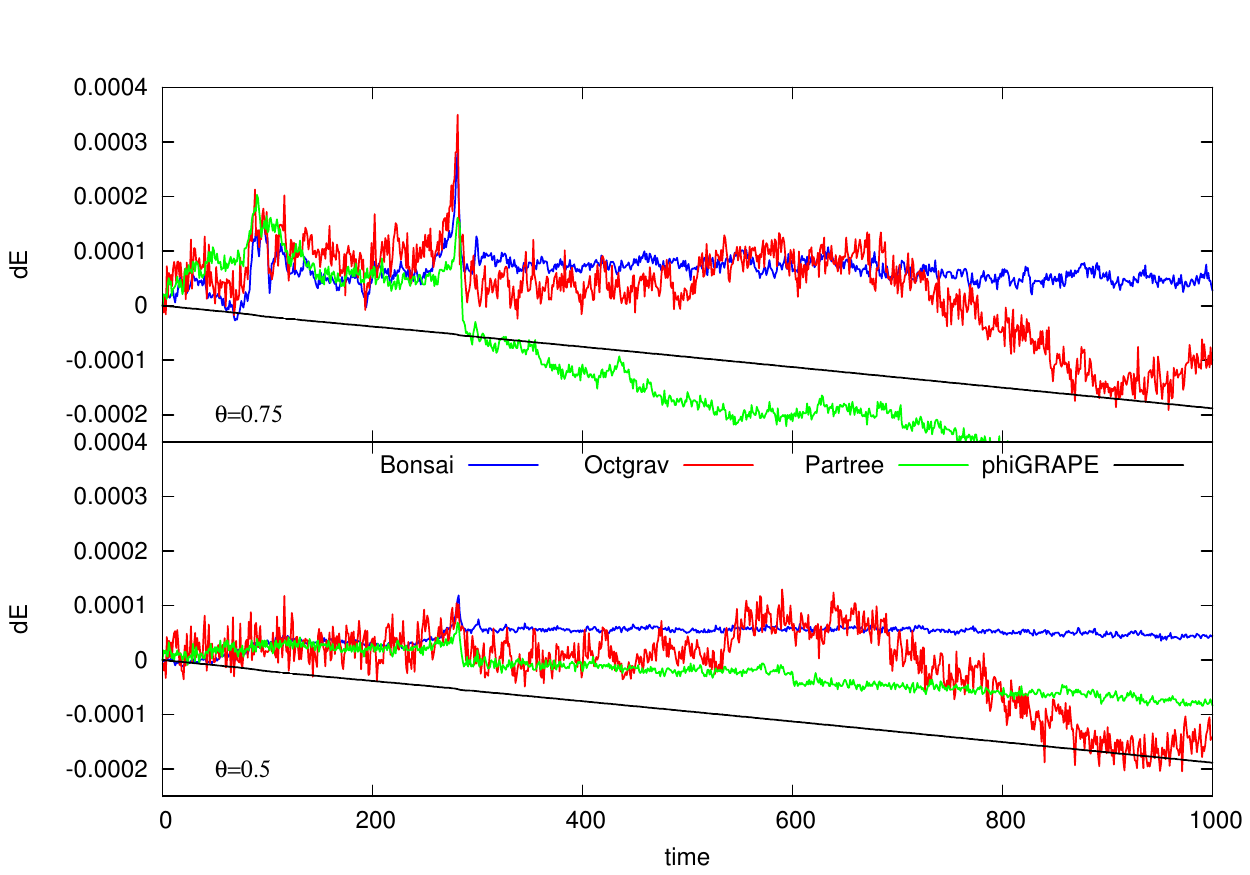}
\caption{Relative energy error over the course of the simulation. The top panel shows 
the results of $\theta=0.75$ and the bottom panel the results of $\theta=0.5$, all other
settings as defined in Tab.~\ref{Tab:MergerProps}. The {\tt Bonsai} results are
shown by the blue lines, the {\tt Octgrav} results are shown by the red lines,
the {\tt Partree} results are shown by the green lines and the black lines show
the results of {\tt phiGRAPE}. Note that the result of {\tt phiGRAPE} is the same in 
the top and bottom panel and that the range of the y-axis is the same in both panels.
}
\label{Image:MergComp1}
\end{figure}

A detailed overview of the energy error is presented in
Fig.~\ref{Image:MergComp1} which shows the relative energy error
($dE$) over the course of the simulation.  Comparing the $dE$ of the
tree-codes shows that {\tt Bonsai} has a more stable evolution than
{\tt Octgrav} and {\tt Partree}. Furthermore if we compare the results
of $\theta=0.75$ and $\theta=0.5$ there hardly is any improvement
visible for {\tt Octgrav} while {\tt Bonsai} and {\tt Partree} show an
energy error with smaller per time-step variance of the energy error. 

The last thing to look at is the execution time of the various codes, 
which can be found in the last column of Tab.~\ref{Tab:MergerProps}. 
Comparing {\tt Bonsai} with {\tt Octgrav} shows that the former is
faster by a factor of $1.6$ and $1.26$ for $\theta=0.75$ and 
$\theta=0.5$ respectively. The smaller speed-up for $\theta=0.5$ 
results from the fact that the tree-traverse, which takes up most
of the execution time, is faster in {\tt Octgrav} than in {\tt Bonsai}. 
Comparing the execution time of {\tt Bonsai} with that of 
{\tt Partree} shows that the former is faster by a factor of 17 (24)
with $\theta=0.75$ ($\theta=0.5$). Note that this speed-up 
is smaller than reported in Section~\ref{Sect:PerfRes} due to 
different initial conditions. 
Finally, when comparing {\tt phiGRAPE} with {\tt Bonsai}, 
we find that {\tt Bonsai} completes the simulation on 1 GTX480 
faster than {\tt phiGRAPE}, which uses 4 GTX480 GPUs,
by a factor of 8.7 ($\theta=0.75$) and 4.9 ($\theta=0.5$).

\section{Discussion and Conclusions}\label{sect:discussion}

We have presented algorithms to efficiently construct and traverse
hierarchical data-structures. These algorithms are implemented as part
of a gravitational $N$-body tree-code. In contrast to other existing
GPU tree-codes, this implementation is executed on the
GPU. While the code is written in CUDA, the methods themselves are
portable to other massively parallel architectures, provided that
parallel scan-algorithms exist for such architectures.  For this
implementation a custom CUDA API wrapper is used that can be replaced
with an OpenCL version. As such the code can be ported to OpenCL, by
only rewriting the GPU functions, which is currently work in progress.

The number of particles processed per unit time, is $2.8$ million 
particles per second with
$\theta=0.75$ on a GTX480. Combined with the stable energy evolution
and efficient scaling permits us to routinely carry out simulations on
the GPU. Since the current version can only use 1 GPU, the limitation
is the amount of memory. For 5 million particles $\pm 1$ gigabyte of
GPU memory is required.

Although the the tree-traverse in {\tt Octgrav} is $\pm10\%$ faster
than in {\tt Bonsai}, the latter is much more appropriate for large
($N > 10^6$) simulations and simulations which employ block-time
steps. In {\tt Octgrav} the complete tree-structure, particle array
and multipole moments are send to the GPU during each time-step. When
using shared-time steps this is a non-critical amount of overhead
since the overall performance is dominated by the tree-traverse which
takes up more than 90\% of the total compute time. However, this
balance changes if one uses block time-steps. The tree-traverse time
is reduced by a factor $N_{\rm groups}/N_{\rm active}$, where 
$N_{\rm active}$ is the number of groups with particles that have to
be updated. This number can be as small as a few percent of $N_{\rm groups}$, 
and therefore tree-construction, particle prediction and communication 
becomes the bottleneck.  By shifting these computations to the
GPU, this ceases to be a problem, and the required host communication
is removed entirely.

Even though the sorting, moving and tree-construction parts of the
code take up roughly 10\% of the execution time,
these methods do not have to be executed during each
time-step when using the block time-step method.  It is sufficient to
only recompute the multipole moments of tree-cells that have updated
child particles. Only when the tree-traverse shows a considerable
decline in performance the complete tree-structure has to be rebuild.
This decline is the result of inefficient memory reads and an increase
of the average number of particle-cell and particle-particle
interactions. This quantity increases because the tree-cell size ($l$)
increases, which causes more cells to be opened by the multipole
acceptance criterion (Eq.~\ref{Eq:Opening}).

Although the algorithms described herein are designed for a
shared-memory architecture, they can be used to construct and traverse
tree-structures on parallel GPU clusters using the methods described
in \cite{169640, 1996NewA....1..133D}.  Furthermore, in case of a
parallel GPU tree-code, the CPU can exchange particles with the other
nodes, while the GPU is traversing the tree-structure of the local
data.  In this way, it is possible to hide most of the communication
time.  

The presented tree-construction and tree-traverse algorithms are not
limited to the evaluation of gravitational forces, but can be applied
to a variety of problems, such as neighbour search, clump finding
algorithms, fast multipole method and ray tracing. In particular, it
is straightforward to implement Smoothed Particle Hydrodynamics in
such a code, therefore having a self-gravitating particle based
hydrodynamics code implemented on the GPU.

\section*{Acknowledgements}

This work is supported by NOVA and NWO grants (\#639.073.803,
\#643.000.802, and \#614.061.608, VICI \#643.200.503, VIDI
\#639.042.607). The authors would like to thank Massimiliano Fatica
and Mark Harris of NVIDIA for the help with getting the code to run on
the Fermi architecture, Bernadetta Devecchi for her help with the
galaxy merger simulation and Dan Caputo for his comments which improved
the manuscript.

\bibliographystyle{elsarticle-num-names.bst}
\bibliography{GBPZ2010} 

\appendix

\section{Scan algorithms}\label{Sect:Scan}

Both tree-construction and tree-traverse make extensive use of parallel-scan algorithms, 
 also known as parallel prefix-sum algorithms. These algorithms are examples of
computations that seem inherently sequential, but for which an efficient parallel
algorithm can be defined. Blelloch \cite{BlellochTR90} defines the scan
operations as follows:
\newline
\newline
{\bf Definition:} \emph{The} all-prefix-sums \emph{operation takes a binary associative
  operator $\oplus$, and an array of n elements}

\begin{center}
[$a_0$, $a_1$, ..., $a_{n-1}$],
\end{center}
\emph{and returns the ordered set}
\begin{center}
[$a_0$,($a_0$ $\oplus$ $a_1$), ...,($a_0$ $\oplus$ $a_1$ $\oplus$ 
... $\oplus$ $a_{n-1}$)].
\end{center}
{\bf Example:} If $\oplus$ is the addition operator, then the all-prefix-sums operation on
the array
\begin{center}
[3 1 7 0 4 1 6 3],
\end{center}
would return
\begin{center}
[3 4 11 11 15 16 22 25].
\end{center}

\noindent 
The prefix-sum algorithms form the building blocks for a variety of methods, including
stream compaction, stream splitting, sorting, regular expressions, tree-depth determination 
and histogram construction. In the following paragraphs, we give a concise account on the
algorithms we used in this work, however we refer the interested readers to the survey by
Blelloch \cite{BlellochTR90} for further examples and detailed descriptions.

\subsection{Stream Compaction}

\begin{figure}
\center
\includegraphics[width=\columnwidth]{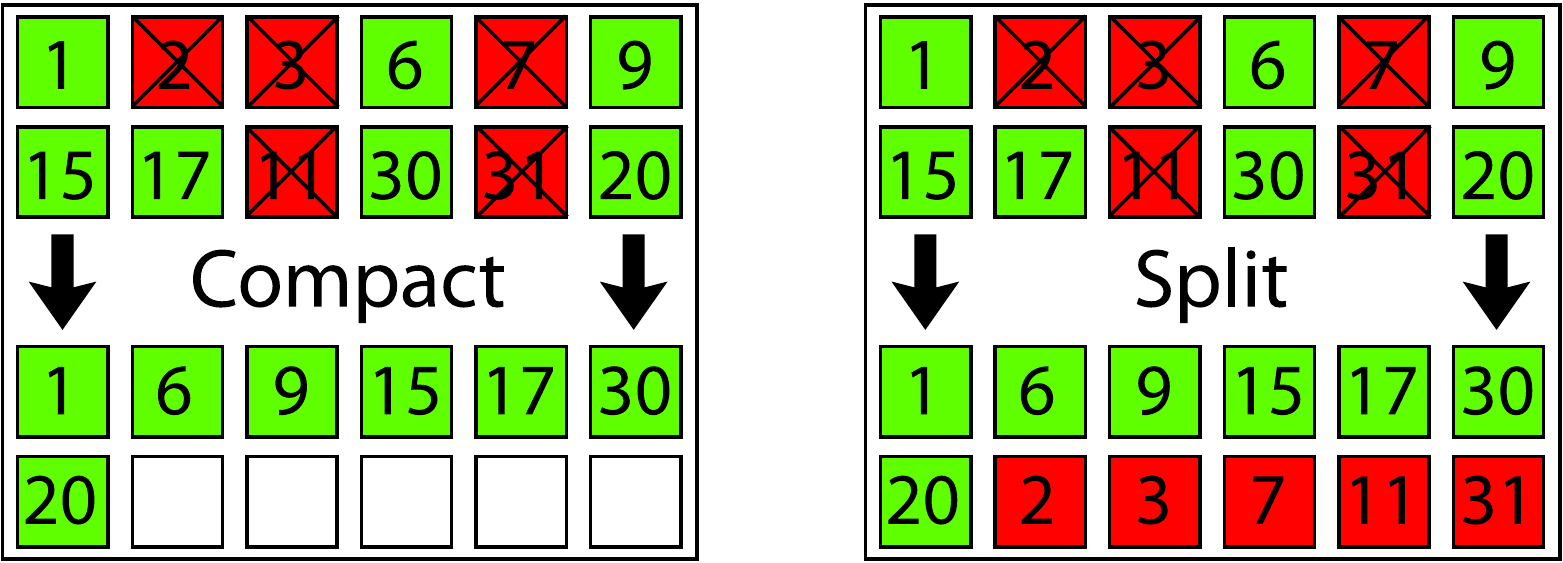}
\caption{Example of compact and split algorithms. The ``compact'' discards invalid items
  whereas the ``split'' places these behind the valid items.  We use this ``split'' primitive
  for our radix sort implementation because it preserves the item ordering--a property which is
  fundamental for the radix sort algorithm.}
\label{Image:ScanExample}
\end{figure}

Stream compaction removes ``invalid'' items from a stream of elements; this algorithm is 
also known as stream reduction.  In the left panel of the Fig.\ref{Image:ScanExample} an
example of a compaction is shown where invalid elements are removed and valid elements 
are placed at the start of the output stream.

\subsection{Split and Sort}

Stream split is related to stream compaction, but differs in that the invalid elements are
placed behind the valid ones in the output stream instead of being discarded (right
panel of Fig.\ref{Image:ScanExample}). This algorithm is used as building block for the
radix sort primitive. Namely, for each bit of an integer we call the split algorithm,
 starting from the least significant bit, and terminating with the most significant
bit \cite{KnuthRadix}.

\subsection{Implementation}

All parts of our parallel octree algorithms use scan algorithms in one way or
another. Therefore, it is important that these scan algorithms are implemented in the most
efficient way and do not become the bottleneck of the application. There are many
different implementations of scan algorithms on many-core architectures \cite{NVIDIA.Scan,
  NVIDIA.ScanTech, BilleterScan}.  We use the method of Billeter et
al. \cite{BilleterScan} for stream compaction, split and radix-sort because it appears to
be, at the moment of writing, the fastest and is easily adaptable for our
purposes. 

Briefly, the method consists of three steps:

\begin{enumerate}
\item Count the number of valid elements in the array.
\item Compute output offsets using parallel prefix-sum.
\item Place valid elements at the output offsets calculated in the previous step.
\end{enumerate}

\noindent We used the prefix-sum method described by Sengupta et al. \cite{NVIDIA.ScanTech},
for all such operations in both the tree-construction and tree-traverse parts of the 
implementation.

\section{Morton Key generation}\label{Sect:MKGen}

One of the properties of the Morton key is its direct mapping between coordinates and 
keys. To generate the keys given a set of coordinates one can make use of look-up tables
or generate the keys directly. Since the use of look-up tables is relative inefficient on 
GPUs (because of the many parallel threads that want to access the same memory) we decided
to compute the keys directly.
First we convert the floating point positions into integer positions. This is done by 
shifting the reference frame to the lower left corner of the domain and then multiply 
the new positions by the size of the domain. 
Then we can apply bit-based dilate primitives to compute the Morton key (List.\ref{list:dilate}, 
\cite{Raman:2008:CDI:1345867.1345918}).
This dilate primitive converts the first 10 bits of an integer
to a 30 bit representation, i.e. {\tt 0100111011} $\rightarrow$ {\tt 000 001 000 000 001
  001 001 000 001 001}: {\tt
\begin{lstlisting}[ caption = {\small The GPU code which we use to dilate the first 10-bits of an integer.}, label=list:dilate ]
int dilate(const int value) {
  unsigned int x; 
  x = value & 0x03FF;
  x = ((x << 16) + x) & 0xFF0000FF;
  x = ((x <<  8) + x) & 0x0F00F00F;
  x = ((x <<  4) + x) & 0xC30C30C3;
  x = ((x <<  2) + x) & 0x49249249;
  return x;
}
\end{lstlisting}
}

This dilate primitive is combined with bit-shift and  {\tt OR} operators to get the particles' key. 
In our implementation, we used 60-bit keys, which is sufficient for an octree with the maximal
depth of 20 levels. We store a 60-bit key in two 32-bit integers, each containing 30-bits of the key. 
The maximal depth imposes a limit on the method, but so far we have never run into problems with our simulations. 
This limitation can easily be lifted by either going to 90-bit keys for 30 levels
or by modifying the tree-construction algorithm when we reach deepest levels. This is something we are 
currently investigating. 

\end{document}